# From the double-stranded helix

# to the chiral nematic phase of B-DNA:

# a molecular model

Fabio Tombolato and Alberta Ferrarini

Dipartimento di Scienze Chimiche, Università di Padova,

via Marzolo 1, 35131 Padova, Italy

B-DNA solutions of suitable concentration form left-handed chiral nematic phases (cholesterics). Such phases have also been observed in solutions of other stiff or semiflexible chiral polymers; magnitude and handedness of the cholesteric pitch are uniquely related to the molecular features. In this work we present a theoretical method and a numerical procedure which, starting from the structure of polyelectrolytes, lead to the prediction of the cholesteric pitch. Molecular expressions for the free energy of the system are obtained on the basis of steric and electrostatic interactions between polymers; the former are described in terms of excluded volume, while a mean field approximation is used for the latter. Calculations have been performed for 130 bp fragments of B-DNA. The theoretical predictions provide an explanation for the experimental behavior, by showing the counteracting role played by shape and charge chirality of the molecule.

# I. Introduction

Solutions of B-DNA display a rich polymorphism as a function of concentration, which comprises the formation of chiral phases.[1] Many aspects of such behavior are still unexplained, as well as unexplained are most phenomena involving DNA-DNA interactions:[2,3,4] the most striking and biologically significant effect is the dense packing of DNA in chromatin. This paper deals with the chiral nematic (cholesteric) phase, which has been observed in vivo and in vitro.[5,6,7,8,9]

Cholesteric phases are also formed by solutions of other stiff or semiflexible chiral polyelectrolytes, e.g. filamentous viruses[10], polysaccharides like xanthan,[11] or columnar aggregates of deoxyguanosine tetramers (G-wires).[9] The latter can have different organizations, depending on the chemical structure of the monomeric units; accordingly, cholesteric phases with different features are formed. In all cases the relation between molecular and macroscopic chirality is unknown, and any simple correlation (molecular helix ↔ phase helicity) fails. An example can illustrate this point. B-DNA helices are right-handed and form a left-handed cholesteric phase, whose pitch increases with temperature. G-wires constituted by the oligodeoxyguanylates denoted as d(GpGpGpG), or $dG_4$, self-assemble in right-handed columns, like B-DNA; however, unlike B-DNA, they form a right-handed cholesteric phase with a pitch which decreases with temperature.[9] Yet the two molecular systems have similar chemical structure and their chiral nematic phases are formed under analogous conditions.

The stability of the chiral nematic phase is explained by the elastic continuum theory in terms of the competition between chiral forces, which favor a twist deformation of the director, and restoring elastic torques, which oppose the deformation.[12] The two contributions reflect the intermolecular interactions specific of each system. The elastic term has a weaker dependence on molecular structure, and reasonable estimates, at least of its order of magnitude, can be obtained even in simple ways. On the contrary, magnitude and sign of the chiral contribution can change dramatically, even for small structural variations; therefore predictions require a detailed modeling of the system. This is a general feature of chiral properties, which are determined by a tiny fraction of the whole intermolecular interactions, with a subtle dependence on the molecular structure.[13] Nevertheless, molecular theories[14,15] and simulation methods[16] able to provide reliable predictions of the cholesteric pitch of thermotropic cholesterics on the basis of the structure of the chiral molecules have been developed. The case of polyelectrolye solutions is more complicated, because these are multicomponent systems characterized by a variety of interactions covering a wide range of lenghtscales. The early theory by Straley deals with a system of hard helices, whose free energy is calculated according to the virial expansion truncated at the second term.[17] The organization of the cholesteric phase is then determined by the best packing of helices. For instance, B-DNA should form a right-handed cholesteric phase, with a pitch independent of temperature, in full contrast with experiment. Subsequently, the additional effect of dispersion interactions between macromolecules in a solvent, treated as a dielectric continuum, was introduced.[18] However, a theory for polyelectrolytes cannot ignore interactions between charges. For example, in the case of DNA each phosphate group has a net charge of -1e at pH~6-7, which corresponds to a high surface charge density of about -1e/nm$^2$. Recently a thorough analysis of electrostatic interactions between stiff polyelectrolytes has been presented, based on a detailed description of screened electrostatic interactions between cylinders with arbitrary surface charge patterns, embedded in a solvent containing microions.[19] This theory provides a deep insight into the effects of the charge distribution on interactions between charged helices. Information about the preferred twist between B-DNA helices can be inferred from the analysis of the potential energy surface as a function of the inter-helix angle and distance.[20]

In this paper we present a molecular theory for the cholesteric pitch of solutions of stiff polyelectrolytes, which differs from previous treatments for two main reasons. A statistical-mechanics analysis is performed, which allows a consistent description of the system properties, comprising orientational order, elastic constants and chiral strength, and of their temperature dependence. Moreover, both steric and electrostatic interactions are considered, according to the same picture which underlies theories developed for the nematic phase of solutions of rod-like polyelectrolytes, following the Onsager suggestion.[21] Such theories are able to explain the orientational order and the features of the isotropic-nematic transition,[22] as well as the elastic properties[23] of cholesteric solutions of stiff chiral polyelectrolytes, approximated as uniformly charged rods. However, the chirality of shape and charge distribution cannot be neglected if chiral properties are investigated. The smallness of chiral effects and their dependence on molecular details pose serious problems to the possibility of getting reliable predictions. On one side an accurate representation of the molecular structure is required, but on the other side a detailed description, e.g. at the atomistic level, is unfeasible for the complex systems under examination. We have considered the case of B-DNA solutions, by taking into account the chirality of shape and charge distribution by a coarse-grained representation, which however preserves the main features of the molecular structure.

The paper is organized as follows. In the next section the theoretical framework is presented and general expressions for the free energy of the system and the cholesteric pitch in terms of steric and electrostatic intermolecular interactions are reported. Then expressions suitable for numerical calculations, in terms of orientational order parameters, are derived, through expansion of the density function on a set of basis functions. In the fourth section the modeling of the B-DNA structure and of intermolecular interactions is described, while in the fifth section the computational aspects of the method are summarized. Then the results obtained are reported and discussed, and conclusive remarks are presented, by pointing out achievements and limits of this work. To avoid making the text unnecessarily heavy, derivations have been confined to Appendices.

## II. Theory

The twist deformation characterizing a chiral nematic phase is specified by sign and magnitude of the pitch of the helix formed in space by the mesophase director $\hat{n}$. The pitch $p$, or correspondingly the wavenumber $q = 2\pi/p$, is defined as positive or negative, according to the right- or left-handedness of the cholesteric helix, respectively.

The continuum elastic theory can justify the stability of the cholesteric phase and provides an expression for the equilibrium pitch in terms of macroscopic properties of the system. Let us consider the Helmholtz free energy density $f$, which is related to the free energy $F$ of the system by the integral

$$F = \int d\mathbf{R}\, f(\mathbf{R}) \tag{1}$$

For small deformations, the expansion of the Helmholtz free energy density $f$ in a power series of the deformation $q$ can be truncated at the quadratic contribution:

$$f \approx f^u + K_t q + \frac{1}{2} K_{22} q^2 \tag{2}$$

where $f^u$ is the free energy density of the undeformed nematic phase, and

$$K_t = \left(\frac{\partial f}{\partial q}\right)_{q=0} \tag{3a}$$

$$K_{22} = \left(\frac{\partial^2 f}{\partial q^2}\right)_{q=0} \tag{3b}$$

are respectively the chiral strength and the twist elastic constant.[12] At given ($T,V$) values, equilibrium corresponds to the minimum of the Helmholtz free energy; by imposing the condition $\left(\frac{\partial f}{\partial q}\right)_{T,V} = 0$, the equilibrium wavenumber is obtained:

$$q = -\frac{K_t}{K_{22}} \qquad (4)$$

The cholesteric pitch vanishes if the mesophase is formed by achiral molecules or by a racemic mixture (with no enantiomeric excess), in which cases $K_t=0$; it has opposite handedness for pairs of mesophases formed by enantiomers, because these have opposite $K_t$ values. The purpose of a molecular theory is the derivation of expressions for chiral strength and twist elastic constant in terms of the intermolecular interactions.

## A. Free energy of the undeformed nematic phase

Let us take a system of $N$ identical particles in the volume $V$ at temperature $T$, interacting through hard core repulsions and electrostatic interactions. Manageable expressions for the Helmoltz free energy of the system are obtained by using a variational approach, based on the Gibbs-Bogoliubov inequality:[24]

$$F \leq F_0 + \langle H_1 \rangle_0 \qquad (5)$$

where $F_0$ is the Helmholtz free energy of a model system with Hamiltonian $H_0$, and $H_1=H-H_0$ is the difference between the Hamiltonian of the system and that of the model. The angular brackets with the zero subscript denote the average value, calculated over the distribution function of the model system:

$$\langle H_1 \rangle_0 = \frac{\int d\Gamma^N (H-H_0)\exp(-H_0/k_B T)}{\int d\Gamma^N \exp(-H_0/k_B T)}. \qquad (6)$$

In this expression $k_B$ is the Boltzmann constant and $\mathbf{\Gamma}^N$ is a collective notation for all the degrees of freedom of the system, $\mathbf{\Gamma}^N=(\mathbf{R}^N, \Omega^N)$, with the vector $\mathbf{R}$ and the Euler angles $\Omega=(\alpha,\beta,\gamma)$ specifying position and orientation of a particle.

We shall now derive expressions for the two contributions in Eq. (5), starting from the case of a uniform system. A reasonable choice for the model is represented by a system hard particles, whose free energy, according to the second virial approximation, is given by [21]

$$F_0 = Nk_BT\left[\ln\left(\Lambda_{tr}^3 \Theta_{or}\right)-1\right] + k_BTV\int d\Omega\, \rho(\Omega)\ln\rho(\Omega) + \qquad (7)$$
$$+ \frac{k_BTV}{2}\int d\Omega_A d\Omega_B \rho(\Omega_A)\rho(\Omega_B) v_{excl}(\Omega_{AB})$$

The first two terms are ideal gas contributions; $\Lambda_{tr} = (h^2/2\pi k_B T m)^{1/2}$ is the de Broglie wavelength and $\Theta_{or} = (h^2/2\pi k_B T I_x)^{1/2}(h^2/2\pi k_B T I_y)^{1/2}(h^2/2\pi k_B T I_z)^{1/2}$, with $h$, $m$ and $I_{x,y,z}$ representing the Planck constant, the mass and the principal components of the inertia tensor.[25] The third term in Eq. (7) is the second virial contribution, with the function $v_{excl}(\Omega_{AB})$ representing the volume excluded to the B by the A particle; this is defined as

$$v_{excl}(\Omega_{AB}) = -\int d\mathbf{R}_{AB}\, e_{AB}(\mathbf{R}_{AB}, \Omega_{AB}) \qquad (8)$$

where the vector $\mathbf{R}_{AB}$ is defined as $\mathbf{R}_{AB}=\mathbf{R}_B-\mathbf{R}_A$, while $\Omega_{AB}$ are the Euler angles for the rotation from the molecular frame of particle A to that of particle B (see Fig. 1). The function $e_{AB}$ is the Mayer function [25]

$$e_{AB}(\Omega_{AB}) = \exp\{-U_h(\mathbf{R}_{AB},\Omega_{AB})/k_BT\} - 1, \qquad (9)$$

with $U_h(\mathbf{R}_{AB},\Omega_{AB})$ the hard core potential between the A and B particles:

$$U_h(\mathbf{R}_{AB},\Omega_{AB}) = \begin{cases} \infty & \text{if A,B overlap} \\ 0 & \text{if A,B do not overlap} \end{cases} \qquad (10)$$

The function $\rho(\Omega,\mathbf{R})$ is the single particle density function,[26] which satisfies the normalization condition:

$$\int d\mathbf{R}\, d\Omega\, \rho(\mathbf{R},\Omega) = N \qquad (11)$$

The density function reflects the particle and phase symmetry properties. In a uniform system the density function is independent of the molecular position; if furthermore the phase is isotropic, it is also independent of orientation, and is simply given by $\rho_{iso} = \frac{1}{8\pi^2 v}$, with $v = \frac{V}{N}$ the available volume per molecule.

Electrostatic interaction between particles are introduced in a mean field way, according to Eq.(6). By taking into account the pairwise additivity of electrostatic interactions and the independence of position of the density function, we can write for a homogeneous system:

$$\langle H_1 \rangle_0 \approx \frac{1}{2} \int d\mathbf{R}_A d\mathbf{R}_B d\Omega_A d\Omega_B \rho(\Omega_A) \rho(\Omega_B) g_h(\mathbf{R}_{AB},\Omega_{AB}) U_{el}(\mathbf{R}_{AB},\Omega_{AB}) \quad (12)$$

with the hard particle pair distribution function $g_h(\mathbf{R}_{AB},\Omega_{AB})$ approximated as

$$g_h(\mathbf{R}_{AB},\Omega_{AB}) = \begin{cases} 0 & \text{if A, B overlap} \\ 1 & \text{if A, B do not overlap} \end{cases} \quad (13)$$

By collecting the contributions in Eqs. (7) and (12) we can write the free energy density of the system as

$$f = f^{id} + f^{ex} \quad (14)$$

with $f^{id}$ representing the ideal term

$$f^{id} = \frac{N}{V} k_B T \left[ \ln(\Lambda_{tr}^3 \Theta_{or}) - 1 \right] + k_B T \int d\Omega \rho(\Omega) \ln \rho(\Omega) \quad (15)$$

and the excess contribution

$$f^{ex} = \frac{1}{2} \int d\mathbf{R}_B d\Omega_A d\Omega_B \, \rho(\Omega_A) \rho(\Omega_B) u(\mathbf{R}_{AB},\Omega_{AB}) \quad (16)$$

where the pair interaction is defined as

$$u(\mathbf{R}_{AB},\Omega_{AB}) = \{-k_B T e_{AB}(\mathbf{R}_{AB},\Omega_{AB}) + g_h(\mathbf{R}_{AB},\Omega_{AB}) U_{el}(\mathbf{R}_{AB},\Omega_{AB})\} \quad (17)$$

We have found an expression for the approximate free energy $F_{app} = F_0 + \langle H_1 \rangle_0$ as a functional of the density function ρ, $F_{app}[\rho]$; according to the Gibbs-Bogoliubov inequality the best approximation to the free energy of the system can be found by functional minimization.

**B. Free energy of the twisted nematic phase and cholesteric pitch**

In the twisted nematic phase the density function depends also on the particle position, and we can write $\rho = \rho(\Omega, \chi(\mathbf{R}))$, where χ is the angle between the local director at the particle position **R** and the director at the origin of the laboratory frame, as shown in Fig. 1. The lengthscale of the twist deformation is orders of magnitude larger than the molecular dimension and the lengthscale of intermolecular interactions. As a consequence, the local phase properties are hardly distinguishable from those of the corresponding nematic phase, and the density function in the cholesteric phase can be reasonably assumed to be the same as that in the undeformed nematic phase, with respect to the local director $\hat{n}$.[27]

Thus, the free energy density of the cholesteric phase can be expressed as the sum of an ideal contribution with the form of Eq. (15), with the Euler angles $\Omega$ expressing the molecular orientation with respect to the local director, along with an excess term which has the form:

$$f^{ex} = \frac{1}{2}\int d\mathbf{R}_B d\Omega_A d\Omega_B \, \rho(\Omega_A)\rho(\Omega_B,\chi(\mathbf{R}_B))u(\mathbf{R}_{AB},\Omega_{AB}) \tag{19}$$

having chosen a laboratory frame with the $Z$ axis parallel to the local director at the position of the A particle.

Then, by reminding the free energy expansion Eq. (2), we obtain molecular expressions for the chiral strength:

$$K_t = \left(\frac{df^{ex}}{dq}\right)_{q=0} = \frac{1}{2}\int d\mathbf{R}_B d\Omega_A d\Omega_B \rho(\Omega_A)\left(\frac{d\rho(\Omega_B,\chi(\mathbf{R}_B))}{dq}\right)_{q=0} u(\mathbf{R}_{AB},\Omega_{AB}) \tag{20}$$

and the twist elastic constant:

$$K_{22} = \left(\frac{d^2 f^{ex}}{dq^2}\right)_{q=0} = \frac{1}{4}\int d\mathbf{R}_B d\Omega_A d\Omega_B \rho(\Omega_A)\left(\frac{d^2\rho(\Omega_B,\chi(\mathbf{R}_B))}{dq^2}\right)_{q=0} u(\mathbf{R}_{AB},\Omega_{AB}) \tag{21}$$

The helical wavenumber can then be obtained as the ratio between chiral strength and twist elastic constant, according to Eq. (4). By distinguishing steric and electrostatic contribution and recalling the form of the pair potential Eq.(17), we can write

$$q = -\frac{\tilde{K}_t^h + \dfrac{K_t^{el}(T)}{k_B T}}{\tilde{K}_{22}^h + \dfrac{K_{22}^{el}(T)}{k_B T}} \tag{22}$$

where the apices *el* or *h* denote electrostatic and hard core contribution, respectively, and $\tilde{K}^h = K^h/k_B T$ is independent of temperature.

## III. Free energy in terms of order parameters

A convenient route for the numerical solution of the problem rests on the technique of expansion of the density function on a basis of orthogonal functions. In this case the appropriate functions are the Wigner rotation matrices $D_{lm}^j(\Omega)$.[28]

By taking into account the $D_{\infty h}$ symmetry of the undeformed nematic phase and approximating particles as axially symmetric, we can write:

$$\rho(\Omega) = \frac{1}{8\pi^2 v} \sum_{j=0,2,4...} (2j+1) \langle D_{00}^j \rangle D_{00}^j(\Omega) \tag{23}$$

with the expansion coefficients $\langle D_{00}^j \rangle$ defined as

$$\langle D_{00}^j \rangle = v \int d\Omega \rho(\Omega) D_{00}^j(\Omega) \tag{24}$$

Such coefficients are the nematic order parameters.

For the density function in the chiral nematic phase we can write:

$$\rho(\Omega, \chi(\mathbf{R})) = \frac{1}{8\pi^2 v} \sum_{j=0,1,2,...} (2j+1) \sum_{l=-j}^{j} \langle D_{l0}^j(\chi(\mathbf{R})) \rangle D_{l0}^j(\Omega) \tag{25}$$

with the coefficients $\langle D_{l0}^j(\chi(\mathbf{R})) \rangle$ defined by an expression analogous to Eq. (24). By using the transformations reported in Fig. 1 and exploiting the addition theorem of Wigner matrices [28] we can write

$$\langle D_{l0}^j(\chi(\mathbf{R})) \rangle = v \sum_{n=-j}^{j} D_{ln}^j(0, \chi, 0) \int d\Omega'' \rho(\Omega'') D_{n0}^j(\Omega'') \tag{26}$$

where the Euler angles $\Omega''$ specify the molecular orientation with respect to the local director in $\mathbf{R}$.

By using the relation $D_{ln}^j(0,\chi,0) = d_{ln}^j(\chi)$, where $d_{ln}^j(\chi)$ are reduced Wigner rotation matrices,[28] and recognizing that the integrals in Eq. (26) are the nematic order parameters, Eq. (24), we can write:

$$\langle D_{l0}^j(\chi(\mathbf{R})) \rangle = d_{l0}^j(\chi) \langle D_{00}^j \rangle \quad j \text{ even} \tag{27}$$

The density function in the twisted nematic phase can then be expressed as

$$\rho(\Omega, \chi(\mathbf{R})) = \frac{1}{8\pi^2 v} \sum_{j=0,2,4,...} (2j+1) \langle D_{00}^j \rangle \sum_{l=-j}^{j} d_{l0}^j(\chi(\mathbf{R})) D_{l0}^j(\Omega) \tag{28}$$

If a laboratory frame with the *Y* axis along the cholesteric axis is chosen, as in Figs. 1 and 2, we can write

$$\chi(\mathbf{R}) = qY \tag{29}$$

where $q$ is the wavenumber of the twist deformation; therefore $\rho = \rho(\Omega, qY)$, and explicit expressions for the derivatives in Eqs. (20) and (21) can be obtained (see appendix A for the derivation):

$$\left(\frac{\partial \rho(\Omega_B, \chi(\mathbf{R}))}{\partial q}\right)_{q=0} = -\frac{1}{16\pi^2 v} \sum_{j=2,4,\ldots} (2j+1)\sqrt{\frac{(j+1)!}{(j-1)!}} \langle D_{00}^j \rangle \times$$

$$\times \left[D_{10}^j(\Omega) - D_{-10}^j(\Omega)\right] Y \tag{30a}$$

$$\left(\frac{\partial^2 \rho(\Omega_B, \chi(\mathbf{R}))}{\partial q^2}\right)_{q=0} = \frac{1}{32\pi^2 v} \sum_{j=2,4,\ldots} (2j+1) \langle D_{00}^j \rangle \times$$

$$\times \left\{-2j(j+1) D_{00}^j(\Omega) + \sqrt{\frac{(j+2)!}{(j-2)!}} \left[D_{20}^j(\Omega) + D_{-20}^j(\Omega)\right]\right\} Y^2 \tag{30b}$$

By introducing Eqs. (23) and (30a) into Eq. (20) for the chiral strength, the following expression is obtained (for the derivation see Appendix C):

$$K_t = \frac{1}{3\sqrt{2}} \frac{1}{8\pi^2 v^2} \sum_{j_A=2,4,\ldots} (2j_A+1) \langle D_{00}^{j_A} \rangle \times$$

$$\times \sum_{j_B=2,4,\ldots} (2j_B+1) \sqrt{\frac{(j_B+1)!}{(j_B-1)!}} C^2(j_A, j_B, 1; 0, 1, 1) \langle D_{00}^{j_B} \rangle \times$$

$$\times \left\{\int d\Omega_{AB} \int d\mathbf{R}_{AB} \, \mathrm{Im}\{D_{10}^{j_B}(\Omega_{AB}) T_{AB}^{11}\} u(\mathbf{R}_{AB}, \Omega_{AB})\right\} \tag{31}$$

Analogously, as shown in Appendix D, by substituting Eqs. (23) and (30b) into Eq. (21) we can express the twist elastic constant as

$$K_{22} = [K_{22}]_{T^{00}} + [K_{22}]_{T^{20}} + \sum_{p=1,2} [K_{22}]_{T^{2p}} \tag{32}$$

with

$$[K_{22}]_{T^{00}} = \frac{1}{8}\left(\frac{1}{8\pi^2 v}\right)^2 \sum_{j_A=0,2,4...}(2j_A+1)\langle D_{00}^{j_A}\rangle \sum_{j_B=2,4,...}(2j_B+1)\langle D_{00}^{j_B}\rangle \delta_{j_A j_B} \times \quad (33a)$$

$$\times\left\{\left(-\frac{16\pi^2}{\sqrt{3}}\right)j_B(j_B+1)C^2(j_A,j_B,0;0,0,0)\right\}\times$$

$$\times \int d\mathbf{R}_{AB}\int d\Omega_{AB} D_{00}^{j_B}(\Omega_{AB}) T_{AB}^{00} u(\mathbf{R}_{AB},\Omega_{AB})$$

$$[K_{22}]_{T^{20}} = \frac{1}{8}\left(\frac{1}{8\pi^2 v}\right)^2 \sum_{j_A=0,2,4...}(2j_A+1)\langle D_{00}^{j_A}\rangle \sum_{j_B=2,4,...}(2j_B+1)\langle D_{00}^{j_B}\rangle \times \quad (33b)$$

$$\times \frac{8\pi^2}{5}\left[j_B(j_B+1)C(j_A,j_B,2;0,0,0)\left(\frac{2}{\sqrt{6}}\right)-\sqrt{\frac{(j_B+2)!}{(j_B-2)!}}C(j_A,j_B,2;0,2,2)\right]\times$$

$$\times C(j_A,j_B,2;0,0,0)\times \int d\mathbf{R}_{AB}\int d\Omega_{AB} D_{00}^{j_B}(\Omega_{AB}) T_{AB}^{20} u(\mathbf{R}_{AB},\Omega_{AB})$$

$$[K_{22}]_{T^{2p}} = \frac{1}{8}\left(\frac{1}{8\pi^2 v}\right)^2 \sum_{j_A=0,2,4...}(2j_A+1)\langle D_{00}^{j_A}\rangle \sum_{j_B=2,4,...}(2j_B+1)\langle D_{00}^{j_B}\rangle \times$$

$$\times \frac{8\pi^2}{5}\left[j_B(j_B+1)C(j_A,j_B,2;0,0,0)\left(\frac{2}{\sqrt{6}}\right)-\sqrt{\frac{(j_B+2)!}{(j_B-2)!}}C(j_A,j_B,2;0,2,2)\right]\times \quad (33c)$$

$$\times 2C(j_a,j_B,2;0,p,p)\times \int d\mathbf{R}_{AB}\int d\Omega_{AB} \operatorname{Re}\{D_{10}^{j_B}(\Omega_{AB}) T_{AB}^{2p}\} u(\mathbf{R}_{AB},\Omega_{AB})$$

$$p=1,2$$

In the above expressions $C(j_A, j_B, j; 0, m, m)$ are Clebsch-Gordan coefficients,[28] while $\mathcal{I}_m\{\}$ and $\mathcal{R}_e\{\}$ denote the imaginary and real part of the function within brackets, respectively. The symbol $T^{1p}$ is used for the irreducible spherical components of the first rank tensor $\mathbf{R}_{AB}$, while $T^{00}$ and $T^{2p}$ are used for the zeroth and second rank irreducible spherical components of the tensor $\mathbf{R}_{AB}\otimes\mathbf{R}_{AB}$ (see Appendix B).

According to Eqs. (31) and (33), calculation of chiral strength and twist elastic constant requires the order parameters $\langle D_{00}^j \rangle$. These can be obtained by minimization of the free energy of the undeformed nematic phase which, by virtue of Eq. (23) can be expressed as a function of the order parameters. Namely, by substituting Eq. (23) into Eqs. (15) and (16), the following expressions for the ideal and excess contribution to the free energy density of the nematic phase, in $k_B T$ unit, are obtained:

$$\frac{f^{id}}{k_B T} = \frac{1}{v}\ln(\Lambda_{tr}^3 \Theta_{or}) - \frac{1}{v}\{1 + \ln(8\pi^2 v)\} + \frac{1}{8\pi^2 v}\sum_{j=0,2,4,\ldots}(2j+1)\langle D_{00}^j \rangle \times$$
$$\times \int d\Omega\, D_{00}^j(\Omega)\left\{\ln \sum_{j'=0,2,4,\ldots}(2j'+1)\langle D_{00}^{j'} \rangle D_{00}^{j'}(\Omega)\right\} \quad (34)$$

$$\frac{f^{ex}}{k_B T} = \frac{1}{16\pi^2 v^2}\sum_{j_A=0,2,4,\ldots}(2j_A+1)\left(\langle D_{00}^{j_A} \rangle\right)^2 \int d\mathbf{R}_{AB}\, d\Omega_{AB}\, D_{00}^{j_A}(\Omega_{AB})\frac{u(\mathbf{R}_{AB}, \Omega_{AB})}{k_B T} \quad (35)$$

## IV. Modeling of B-DNA and Intermolecular Interactions

Calculation of chiral strenght $K_t$, elastic constant $K_{22}$ and order parameters $\langle D_{00}^j \rangle$ requires evaluation of integrals of the general form:

$$\int_0^\infty dR_{AB} R_{AB} \int_0^{2\pi} d\phi_{AB} \int_0^\pi d\vartheta_{AB} \sin\vartheta_{AB} \int_0^{2\pi} d\alpha_{AB} \int_0^{2\pi} d\beta_{AB} \sin\beta_{AB} \int_0^{2\pi} d\gamma_{AB} \times$$
$$\times u(R_{AB}, \phi_{AB}, \vartheta_{AB}, \alpha_{AB}, \beta_{AB}, \gamma_{AB}) \Xi(R_{AB}, \phi_{AB}, \vartheta_{AB}, \alpha_{AB}, \beta_{AB}) \quad (36)$$

where $\mathbf{R}_{AB}$, the vector position of molecule B with respect to molecule A, is expressed in spherical coordinates and $\Xi(R_{AB}, \phi_{AB}, \vartheta_{AB}, \alpha_{AB}, \beta_{AB})$ denotes a generic function, whose specific form depends upon the property we are dealing with. In particular we have

$$\Xi(\mathbf{R}_{AB},\Omega_{AB}) = \begin{cases} \text{Im}\{D_{10}^{j_B}(\Omega_{AB})R_{AB}^1\} & \rightarrow K_t \quad (37a) \\ \text{Re}\{D_{i0}^{j_B}(\Omega_{AB})T_{AB}^i\} & \rightarrow K_{22} \quad (37b) \\ D_{00}^{j_A}(\Omega_{AB}) & \rightarrow \text{order parameters} \quad (37c) \end{cases}$$

The cost of the calculation can be very high since a large number of pair configurations has to be sampled. In particular, high accuracy is required in evaluating $K_t$, because this is a small quantity resulting from the sum of integrals which are large in value and opposite in sign. As will be shown below, each integral requires evaluation of the integrand function for a number of pair configurations of the order of $10^9$. If the molecule is represented as an assembly of $N$ spheres with $M$ point charges, a number of operations proportional to $N^2$ and $M^2$ is required for each configuration. It follows that the feasibility of calculations depends on the level of detail employed in modelling the molecular features.

## A. Modelling the shape of B-DNA

The function in Eq. (37a) estracts the chiral part of intermolecular interactions. Of course steric interactions will have a chiral component only if the two interacting particles have a chiral shape. On the contrary, the presence of a chiral charge distribution is not a sufficient condition for a non-vanishing electrostatic contribution to the chiral strength. Namely, on the basis of symmetry considerations it can be demonstrated that the electrostatic contribution to $K_t$ vanishes for cylinders with a helical charge distribution on their surface (see Appendix E). So a chiral shape is necessary to have not only a steric, but also an electrostatic contribution to $K_t$.

We have modelled double stranded B-DNA in the following way: a base pair is represented by five spheres, one corresponds to the aromatic cores, two to the sugar and two to the phosphate groups. The centres of all spheres lie on the same plane, perpendicular to the helix axis (see Fig. 3). The DNA helix is obtained by a 3.4 Å translation of the centre of the spheres representative of the aromatic cores along a common axis and a 36° right-handed rotation of the base pair plane about such an axis. The geometry parameters are reported in Table I; a fragment of about 30 bp of our model B-DNA is shown in Fig. 4. Despite its simplicity, this model can reproduce the main features of B-DNA, with realistic dimensions of major and minor groove. This model has been proposed by Abascal and Montoro[29] and subsequently used by other authors[30] to simulate the distribution of microions around B-DNA.

For polymers with a strong shape anisotropy, like DNA fragments, the elastic constant $K_{22}$ and the order parameters weekly depend on molecular details. So, with the aim of reducing the computational cost, we have calculated these properties by approximating a DNA molecule as an assembly of fused spheres of radius equal to 11.9 Å, with the centres aligned along a common axis, at a distance of 3.4 Å from each other. Because of its shape this model will be henceforth denoted as the "spherocylinder".

**B. Intermolecular electrostatic potential and charge parameterisation**

Electrostatic interactions between polyelectrolytes are mediated by solvent, counterions and salt; this makes their description a formidable task. According to the linearized Poisson-Boltzmann theory, screened Coulomb interactions between polyelectrolytes can be considered; thus, we can write for a pair of polyelectrolytes, say A and B:

$$U_{el}^{AB}(\mathbf{R}_{AB}, \Omega_{AB}) = \frac{e^2}{4\pi\varepsilon_0\varepsilon} \sum_{i=1}^{M_A} \zeta_{Ai} \sum_{j=1}^{M_B} \zeta_{Bj} \frac{\exp(-\kappa_D |\mathbf{r}_{Ai} - \mathbf{r}_{Bj}|)}{|\mathbf{r}_{Ai} - \mathbf{r}_{Bj}|} \quad (38)$$

where $M_A$ ($M_B$) is the number of point charges in polyelectrolyte A (B), $e\zeta_{Ai}$ ($e\zeta_{Bj}$) is the $i$-th ($j$-th) charge on molecule A (B) and $\mathbf{r}_{Ai}$ ($\mathbf{r}_{Bj}$) is its position, $\varepsilon$ is the dielectric permittivity of the solvent and $\varepsilon_0$ that of vacuum. The parameter $\kappa_D^{-1}$ is the so called Debye screening length:

$$\kappa_D = \left(\frac{2Ie^2}{\varepsilon_0\varepsilon k_B T}\right)^{\frac{1}{2}},$$ with the ionic strength $I = \frac{10^{-3}}{2\mathcal{N}} \sum_\alpha z_\alpha^2 \rho_\alpha^0$ [mol/l]. Here $\mathcal{N}$ is the Avogadro number, $z_\alpha$ is the valence of $\alpha$ ions and $\rho_\alpha^0$ their concentration in bulk solution. Eq. (38), which derives from a mean field treatment of the micro-ion atmosphere, is more satisfactory in the case of monovalent micro-ions, which are characterized by weak correlations; moreover, it is reasonable only for low values of the surface charge density on polyelectrolytes. Actually, at sufficient distance between the polymer it can still be used for higly charged polyelectrolytes, like DNA, provided that effective charges are used.

We have used Eq. (38) with effective charges defined according to the Manning theory.[31] So, the fraction of uncompensated charge is given by $\delta = \frac{1}{|z|\xi}$, with $z$ equal to the counterion valence, and the parameter $\xi$ (>1) defined as $\xi = \frac{e^2}{(4\pi\varepsilon_0\varepsilon b k_B T)}$, where $b$ is the average charge spacing along the helix axis. For DNA, $b = 0.34\text{nm}/2$ (two phosphate groups with charge -1e at distance 0.34 nm); for monovalent counterions in water at 25°C, we obtain $\xi = 4.2$ and $\delta = -0.24$.

As will be emphasised below, most configurations important for the emergence of chiral effects are characterised by the presence of small portions of the two polyectrolytes in close proximity; an example is shown in Fig. 5. The underlying assumptions of Eq. (38) are certainly inappropriate for interactions between charges located in such regions. With the purpose of taking into account this effect, while keeping the simple form necessary in the present framework, we have introduced a phenomenological hybrid potential, defined as follows. The electrostatic interaction between the $i$th charge (belonging to the A molecule) and the $j$th charge (in the B molecule) is assumed to have the Coulomb form if their distance ($r_{ij}$) is equal to the contact distance between the spheres bearing them ($\sigma_{ij}$). For distances larger than a given reference distance ($r_0$), the form of Eq. (38) is assumed, and a linear interpolation between the boundary values is taken for intermediate distances:

$$U_{ij}^{el}(\mathbf{r}) = \begin{cases} \dfrac{e^2 \zeta_{Ai} \zeta_{Bj}}{4\pi\varepsilon_0 \varepsilon' r_{ij}} & r_{ij} = \sigma_{ij} \\ \dfrac{e^2 \zeta_{Ai} \zeta_{Bj}}{4\pi\varepsilon_0 \varepsilon r_{ij}} \exp(-\kappa_D r_{ij}) & r_{ij} \geq r_0 \\ \dfrac{U_{ij}^{el}(r_0) - U_{ij}^{el}(\sigma_{ij})}{r_0 - \sigma_{ij}}(r_{ij} - \sigma_{ij}) & \sigma_{ij} > r_{ij} > r_0 \end{cases} \tag{39}$$

where the symbol $\varepsilon'$ denotes an effective dielectric permittivity.

In our calculations for B-DNA, negative effective charges of magnitude the $\delta e$ have been located at the positions shown in Fig. 3, which correspond to the centres of the spheres representing phosphate groups in the model used for calculation of the chiral strength $K_t$.

## V. Computational Methods

Calculation of the cholesteric pitch at given temperature, concentration and ionic strength requires:
i)      evaluation of integrals of the general form Eq. (36);
ii)     calculation of order parameters.

As explained at the beginning of the previous Section, the first step is computationally demanding and calculations are only feasible if efficient algorithms and optimized procedures are used. Sensible choices have enabled a sizeable reduction of computing time; calculation of the cholesteric pitch can be carried out in about a week on a 2000MHz desktop computer. The strategy adopted will be outlined in the following.

Calculation of the electrostatic contribution is the most time consuming step. It turns out to be convenient to store the electrostatic potential generated by molecule A on a grid; this allows a reduction of the number of computational steps, which in this way scale with the number of point charges in the polymer $M$, rather than with $M^2$. The need of a grid dense enough to ensure a satisfying degree of accuracy requires the storage of a huge array. In the case of the spherocylinder model, further advantage can be derived from the fact that integration on the $\gamma_{AB}$ variable in Eq. (36) can be avoided. Namely, it can be shown that this integration has the effect of transforming the interaction between a point charge on molecule A and all the $M$ discrete charges on molecule B in the interaction between the charge on A and $M/m$ uniformly charged rings on B, where $m$ is the number of charges which are located at the same height and distance from the cylinder axis. In this case each point of the grid represents the electrostatic potential integrated over the $\gamma_{AB}$ angle.

According to Eq. (36), each term requires evaluation of six-fold integrals (five-fold in the case of the spherocylinders). The following integration order is employed:

$$\int_0^\pi d\vartheta_{AB} \sin\vartheta_{AB} \int_0^{2\pi} d\alpha_{AB} \int_0^\pi d\beta_{AB} \sin\beta_{AB} \int_0^{2\pi} d\phi_{AB} \int_0^{2\pi} d\gamma_{AB} \int_0^\infty dR_{AB} R_{AB} \ldots \quad (40)$$

The innermost integral is over the $R_{AB}$ variable; calculation of both electrostatic and steric interactions requires the identification of the closest approach distance $R_{AB}^0$ for all possible pair configurations. This has been accomplished by using the algorithm outlined in Appendix F. The number of checks, thus the computing time, can be significantly reduced in the following way: a first estimate of the closest approach distance is obtained for spherocylinders; once identified the large spheres in contact, a finer evaluation of the contact distance is performed on a restricted region of the helices. Typically, about 40 base pairs above and 40 below that corresponding to the large sphere are taken.

The excluded volume integral is analytical and is calculated between the integration limits 0 and $R_{AB}^0$. On the contrary, the electrostatic contribution has to be evaluated numerically. Because of the regular form of the integrand function, integration has been simply performed by using the trapezoidal rule.[32] In principle, calculations should be performed for $R_{AB}$ ranging from $R_{AB}^0$ to infinity. In practice, the upper integration limit is given a finite value, $R_{AB}^1$. For the calculation of contributions to order parameters and twist elastic constants we have assumed $R_{AB}^1 = R_{AB}^0 + 10\kappa_D^{-1}$, a value large enough to ensure that the integrand is negligible, and 60 integration points have been taken. In the case of contributions to the chiral strength $K_t$, the integration can be truncated at the value $R_{AB}^1$ equal to the closest approach distance between the spherocylinders enclosing the helices. Namely, for larger distances any chiral contribution is excluded for symmetry reasons, as mentioned above. Because of the smallness of the integration range, about some Ångstroms, it is sufficient to use only three integration points; this is a considerable advantage from the point of view of computing time.

The dependence of the integrand on the $\gamma_{AB}$ angle is quite irregular for the helical particles, with sharp maxima, especially in the case of chiral contributions. A convenient quadrature algorithm has been devised in the Romberg method, which allows a non-uniform spacing of abscissas.[32] Integrals on the variables $\vartheta_{AB}, \beta_{AB}, \alpha_{AB}, \phi_{AB}$ have been calculated using the Gauss algorithm, with integration points and weights determined by the zeros of the Legendre (for the $\vartheta_{AB}$ and $\beta_{AB}$ variables) and the Chebyshev polynomials (for the $\alpha_{AB}$ and $\phi_{AB}$ variables).[32] For the $\phi_{AB}$ function, characterized by rapid oscillations between positive and negative values, 192 integration points have been used. On the contrary, for the $\vartheta_{AB}$ and $\beta_{AB}$ variables only 12 points have been taken for calculating $K_t$, while 12 and 24 integration points, respectively, have been used to evaluate order parameters and elastic constant. Only a single value was sufficient for the $\alpha_{AB}$ angle, because of the scarce dependence of the integrals upon this variable.

For given thermodynamic conditions, i.e. given the values of the variables (*N,V,T*), the equilibrium order parameters of the nematic phase are obtained by minimization of the approximate Helmholtz free energy $F_{app}(\langle D_{00}^2 \rangle, \langle D_{00}^4 \rangle, ...)$, as explained in Appendix G. In principle the stability against phase separation should be checked; however we have neglected this possibility, in view of our aim, which is not an accurate description of the features of the phase transition, but rather a reasonable estimate of order parameters, consistent with the choices made for the calculation of elastic constants and chiral strength. The free energy density has been minimized by using the Powell algorithm.[32]

## VI. Results and discussion

Calculations have been performed for an aqueous solution of 130 bp fragments of B-DNA, for which pitch values at different temperatures and ionic strengths are available.[8][9] Two temperatures, 286 K and 323 K, and three different ionic strengths, I=0.1, 0.2 and 0.5 mol/l have been considered. The concentration parameters used for the calculations are reported in Table II. It has to be mentioned that in evaluating the ionic strength also counterions are taken into account; so, in the absence of added salt we have I= 0.1mol/l for a DNA concentration equal to 200 mg/ml. This point has to be taken into account when theoretical predictions are compared with experimental values, because in most cases the concentration of added salt is reported. The parameters required to specify the electrostatic interaction, i.e. the Debye screening constant $\kappa_D$ and the fraction of uncompensated charge $\delta$, are shown in Table III; they have been calculated at different temperatures and ionic strengths according to the expressions reported in Sec. IV.B, by taking into account the temperature dependence of the dielectric constant of the solvent. We can see that, as a result of two compensating temperature effects, $\kappa_D$ is practically the same at $T$=286 K and i=323 K. The dielectric constant $\varepsilon'$ appearing in Eq. (39) has been given the value 2, while the reference distance $r_0$ = 8Å has been assumed. This choice corresponds to a distance of 2Å between the surface of the spheres representing phosphate groups.

We shall start considering a hypotetical DNA solution with purely steric interactions. Of course this is not an appropriate model for a strong polyelectrolyte like DNA, but it will be useful to understand the results obtained for the more realistic model reported henceforth. Table IV displays the predicted order parameters; only values up to the sixth rank have been calculated since, as will be discussed in more detail below, truncation at this level of the summations in Eqs. (31-35) guarantees a good compromise between accuracy and computing speed. We can see that, under the conditions chosen for the calculations, a high order is predicted, with order parameters which decrease with increasing rank, but remain non-negligible even at the sixth rank. Table V reports the twist elastic constant $K_{22}$, the chiral strength $K_t$ and the cholesteric wavenumber $q$. Experimental order parameters and elastic constant are not available, but the values obtained for $\langle D_{00}^2 \rangle$ and $K_{22}$ appear reasonable for nematic solutions of stiff polymers.[23] The chiral strength $K_t$ is negative, this means that steric interactions drive the formation of a right-handed cholesteric helix phase. This result, in agreement with the Straley theory, can be explained by considering that the cholesteric handedness is determined by configurations with molecules fitting into each other grooves (see Fig. 5). So, a right-handed cholesteric phase is predicted for the system of hard helices; the calculated cholesteric pitch, of about 35 μm, is about 15 times longer than the experimental value. Of course no temperature dependence is predicted for the hard particle system, in clear contrast with experiment.

When the electrostatic interactions are switched on, the results reported in Tables VI and VII are obtained. We can see that electrostatic interactions have a very small effect on order parameters and elastic constants, which can be summarized by saying that the mesophase becomes slightly less ordered and more easily deformable. The reason is that electrostatic interactions between equally charged molecules oppose their parallel alignment.[23] On the contrary, electrostatic interactions have a dramatic effect on the cholesteric pitch, since they give a large contribution to the chiral strength $K_t$, opposite in sign to the steric term. This fact can be explained considering that electrostatic repulsions are maximized in those configurations which are favored for sterical reasons, because charges of equal sign lie at close distance. The electrostatic contribution to chiral strength largely overcomes the steric one; therefore a left-handed cholesteric phase is predicted, as experimentally observed. Also magnitude and temperature dependence of the pitch are in agreement with experiment, as appears from Fig. 6, where the theoretical predictions are compared with the experimental data reported in ref. 9 (calculations and measurements refer to constant *V* and constant *p*, respectively, but this difference can be neglected for our system in the temperature range under investigation). The pitch lengthens with increasing temperature (the wavenumber *q* decreases), and the reason can be easily understood with the aid of Eq.(22). We should recall that the excluded volume contribution is independent of *T*. The electrostatic term at the denominator is small; so, as a first approximation, the temperature dependence of the denominator can be neglected. The ratio $K^{el}(T) / k_B T$ at the numerator decreases with increasing temperature. It follows that the relative weight of the steric contribution, which plays against a left-handed distortion of the nematic phase, becomes stronger at higher temperature.

Tables VI and VII also report the values of order parameters, elastic constants, chiral strengths and cholesteric wavenumbers predicted at *T*=323 K and different ionic strengths. We can see that the electrostatic contribution to order parameters and elastic constants decreases with increasing ionic strength; thus, their values get even closer to those obtained for the system of hard helices. Also the electrostatic contribution to $K_t$ decreases with increasing ionic strength, but it remains much larger then the steric contribution; as a result, a pitch variation of the about 30% is predicted on going from I=0.1 and I=0.5 mol/l, a result compatible with experimental data.[8,9]

It is now worth considering the effect of some assumptions and approximations used in our calculations. We shall start from the form of the electrostatic potential, Eq. (39). The introduction of the short- range Coulomb form is only a simple way to correct for the unrealistic screening of electrostatic interactions between regions of polymers at very short distance, and the dielectric constant $\varepsilon'$ and reference distance $r_0$ should be taken as phenomenological parameters. The choice of their values affects the magnitude of the electrostatic contribution to $K_t$ and its dependence on the ionic strength. The electrostatic contribution to the chiral strength at T=323 K and I=0.2 mol/l is equal to $1.50 \times 10^{-6}$ N/m when calculated with $\varepsilon'=2$ and $r_0 = 8$Å, and drops to $0.23 \times 10^{-6}$ N/m if a purely screened Coulomb potential is used. This corresponds to a larger value of the predicted cholesteric pitch. Actually, $\varepsilon'$ and $r_0$ could be taken as adjustable parameters, to be determined from comparison with experimental data. The values obtained in this way would be very close to those we have taken in our calculations, on the basis of the reasonable assumptions that $\varepsilon'$ should be similar to the dielectric constant of an organic medium, while $r_0$ should roughly correspond to a distance comparable with the dimension of a water molecule between the surfaces of two polyelectrolytes.

Another point to be checked is the effect of truncation in the summations appearing in Eqs. (31- 35). The magnitude of terms decreases with increasing values of the indices $j_A$, $j_B$, and the series can be truncated at a finite value, $j^{max}$. This can be estimated by taking into account that: (i) high accuracy is required because the integrands are oscillating functions with a number of zeroes increasing with rank; (ii) integrals are multiplied by coefficients that increase with rank. We have taken $j_A^{max} = j_B^{max} = 6$; the contributions of different rank to twist elastic constant and chiral strength are shown in table VIII and IX. These are defined in such a way that $K_{22} = \sum_{j_A j_B} K_{22}^{j_A j_B}$ and $K_t = \sum_{j_A} K_t^{j_A j_A}$. It appears that even the sixth rank contributions do not vanish; anyway the magnitude of terms decreases with rank, with a rate which depends on the kind of interaction and property. We can see that $K_{22}^{j_A j_B} = K_{22}^{j_B j_A}$; this result, which is in accordance with symmetry requirements, gives an indication on the accuracy of our calculations. The effect of truncation on twist elastic constant, chiral strength and cholesteric wavenumber is shown in Table X. The non-negligible contribution of terms of rank higher than the second appears; namely, truncation at the second rank would give an error of about 100% on the predicted pitch.

## VII. Conclusion

We have presented a molecular theory and a numerical procedure for the prediction of the pitch of chiral nematic solutions of stiff polyelectrolytes on the basis of their structure.

A statistical-mechanics approach is used to connect phase properties and intermolecular interactions. The challenge is represented by the need of combining a coarse grained description, imposed by the complexity of the system, with a sufficiently detailed representation of the chiral molecular properties at the origin of the macroscopic phenomenon.

To this purpose several approximations have been taken. First of all, the intermolecular interactions between polyelectrolytes have simply been described as the superposition of excluded volume and electrostatic contributions, with the solvent regarded as a dielectric continuum and ions viewed as an ionic density which screens interactions between polymer charges. Moreover, charge interactions are simply treated at the level of the linearized Poisson-Boltzmann equation, with the use of renormalized charges, defined according to the Manning theory; a phenomenological correction has been introduced for non-screened interactions between the small portions of polymers which are at very short distance. Effects of dielectric discontinuity have been fully neglected.

Calculations have been performed for 130 bp fragments of B-DNA, by using a representation of the molecule as a regular rigid helix, so neglecting the deviations resulting from its flexibility and from the presence of different base pairs. Relaxing the approximations would certainly affect the numerical results; however the picture emerging from our work should not change, since this depends on the physical ingredients present in the model.

In view of all the underlying approximations, we cannot expect a strict agreement between theoretical predictions and experimental data, but this is not our scope. We rather intend to provide some insight into the mechanisms, up to now unclear,[2] driving the formation of chiral nematic phases in solutions of polyelectrolytes, on the basis of a physically reasonable picture.

The main results of this work can be summarized in the following way.

- A model based on steric and electrostatic interactions between polyelectrolytes correctly predicts the organization of the cholesteric phase of B-DNA solutions. It provides and explanation for the observed left-handed helicity and for the temperature dependence of the pitch. We would like to stress here the importance of handedness in characterizing a cholesteric phase; this point seems sometimes to be undervalued, as though it were obvious. On the contrary, it has been seen experimentally and explained theoretically that the relation between cholesteric handedness and molecular structure is by no way simple, and small structural changes, sometimes occurring even in the same molecules as a consequence of a change of temperature or solvent, can be associated with pitch inversion. [33] [34] [35] Thus, sign is an important piece of information, which should always accompany the magnitude of the cholesteric pitch (although experimental determination might be a non trivial task) and should not be ignored by theories for chiral nematic phases. By assuming reasonable values of the parameters describing electrostatic interactions, we predict for the cholesteric phase of B-DNA a pitch of a few micrometers, which increases with temperature, in agreement with experiment. As pointed out by some authors,[36] knowledge of the temperature dependence of the cholesteric pitch can be extremely useful for understanding the mechanism underlying this phenomenon of chirality amplification.

- Steric and electrostatic interactions are generally recognized as the responsible for order and elastic properties of nematic solutions of stiff polyelectrolytes; [22] [23] [37] they also determine the alignment of proteins in nematic suspensions of filamentous viruses. [38] [39] In both cases theoretical models based on the excluded volume representation of steric repulsions and a mean field description of electrostatic interactions between charged macromolecules have been shown to be able to explain the experimental data. Depending on the property under investigation, a greater or lesser detail is required in the description of molecular shape and charges. A simple representation of cylinders as uniformly charged rods can be sufficient in the former case, while in the latter a realistic account of the protein structure and charge distribution is necessary. For the prediction of the cholesteric pitch, the chirality of both shape and charge pattern have to be taken into account.

- The shape chirality plays a crucial role in our description. Not only it is at the origin of the steric contribution to the cholesteric pitch, but it is also responsible for the emergence of the electrostatic contribution. Namely, as already pointed out by some authors, [40] [41] chiral interactions would be washed out if averaged over a uniaxial distribution. Within our model it is the shape chirality which breaks the axial symmetry of the pair distribution function at short distances. It follows from this issue that chiral effects are determined by those pair configurations in which at least parts of the polymers lie at short distance, so that they can feel each other shape chirality.

- Calculations show that steric and electrostatic interactions have a counteracting effect in the case of B-DNA, which can be simply explained in the following way. Steric interactions would lead to the formation of a right-handed cholesteric phase, as a consequence of the good packing of helices fitting into each other grooves. However such configurations are also characterized by strong repulsions between the charges of polyelectrolytes; therefore they are unfavorable for electrostatic reasons. Since for B-DNA solutions the latter effect prevails, a left-handed cholesteric phase results.

It is experimentally observed that the cholesteric pitch of B-DNA solutions is strongly affected by intercalators, groove binders and polycations like spermidine.[42][43] The presence of molecules interacting with DNA is likely to change the parameters of the molecular helix and influence the effective charges of DNA. So, according to our model a change of steric and electrostatic contributions to chiral strength and twist elastic constant is expected, with strong effects on the cholesteric pitch.

The approach developed in this work is not restricted to the case of DNA, and could be applied to the chiral nematic phase of other polyelectrolytes. In particular it would be interesting to investigate the case of G-wires. On the basis of the results obtained for B-DNA, we can attempt an explanation for behavior of $dG_4$ aggregates mentioned in the Introduction: as for B-DNA, we expect that steric and electrostatic interactions should favor a right-handed and a left-handed cholesteric phase, respectively. However, unlike the case of B-DNA, the former contribution would prevail for $dG_4$. Such a picture should be checked with calculations; a practical difficulty derives form the uncertainty about the aggregate geometry, which has to be known with a certain level of detail for reliable predictions of the cholesteric phase organization.

## Acknowledgements

This work has been supported by the Italian Ministry for University and Research (MIUR) through projects PRIN2003 ("Cristalli Liquidi e Macromolecole per Nanostrutture Organizzate") and FIRB2001 (RBNE01P4JF). The authors gratefully acknowledge prof. Gian Piero Spada and dr. Stefano Moro for stimulating discussions.

# Appendix A

**Derivation of the expression for the q derivatives of the density function, Eqs. (30a-c)**

By using the expansion of the density function Eq. (28), the *q* derivatives in Eqs. (20) and (21) can be expressed as:

$$\left(\frac{\partial \rho(\Omega_B, \chi(\mathbf{R}_B))}{\partial q}\right)_{q=0} = \frac{1}{8\pi^2 v} \sum_{j=0,2,4,\ldots} (2j+1)\langle D_{00}^j \rangle \sum_{l=-j}^{j} \left(\frac{d\, d_{l0}^j(\chi(\mathbf{R}_B))}{d\chi}\right)_{\chi=0} D_{l0}^j(\Omega_B) Y_B \tag{A.1a}$$

$$\left(\frac{\partial^2 \rho(\Omega_B, \chi(\mathbf{R}_B))}{\partial q^2}\right)_{q=0} Y_B^2 = \frac{1}{8\pi^2 v} \sum_{j=0,2,4,\ldots} (2j+1)\langle D_{00}^j \rangle \sum_{l=-j}^{j} \left(\frac{d^2 d_{l0}^j(\chi(\mathbf{R}_B))}{d\chi^2}\right)_{\chi=0} D_{l0}^j(\Omega_B) Y_B^2 \tag{A.1b}$$

So, the $\chi$ derivatives of the reduced Wigner matrices have to be calculated. The following expression has been used:[28]

$$d_{l0}^j(\chi) = \left[(j+l)!(j-l)! j! j!\right]^{\frac{1}{2}} \sum_k (-1)^k \frac{\left(\cos\frac{\chi}{2}\right)^{2j-2k+l} \left(\sin\frac{\chi}{2}\right)^{2k-l}}{k!(j+l-k)!(j-k)!(-l+k)!} \tag{A.2}$$

with the summation extended to all *k* values which allow non negative values for the argument of the factorials.

By taking into account the condition on the factorials and those of vanishing exponents of the sin($\chi$/2) powers in the $\chi$ derivatives of the reduced Wigner matrices, it follows:

$$\frac{d}{d\chi/2} d_{l,0}^j(\chi) = -\frac{1}{2}\sqrt{\frac{(j+1)!}{(j-1)!}}(\delta_{l,1} - \delta_{l,-1}) \tag{A.3a}$$

$$\frac{d^2}{d(\chi/2)^2} d_{l,0}^j(\chi) = \frac{1}{4}\left[-2j(j+1)\delta_{l,0} + \sqrt{\frac{(j+2)!}{(j-2)!}}(\delta_{l,2} + \delta_{l,-2})\right] \tag{A.3b}$$

where $\delta$ is the Kronecker symbol.

## Appendix B

**Expression for $Y_B$ e $Y_B^2$ in terms of irreducible spherical components**

The cartesian component $Y_B$ and its square can be expressed as the following linear combinations [28]

$$Y_B = \left(\frac{1}{2}\right)^{\frac{1}{2}} i\left(T_B^{11} + T_B^{1-1}\right) \tag{B.1a}$$

$$Y_B^2 = -\frac{1}{\sqrt{3}} T_B^{00} - \frac{1}{\sqrt{6}} T_B^{20} - \frac{1}{2} T_B^{22} - \frac{1}{2} T_B^{2-2} \tag{B.1b}$$

where $T^{1p}$ are the irreducible spherical components of the first rank tensor $\mathbf{R}_B$, while $T^{00}$ and $T^{2p}$ are the zeroth and second rank irreducible spherical components of the tensor $\mathbf{R}_B \otimes \mathbf{R}_B$. All tensors in this equation are expressed in the laboratory frame; if instead a reference frame fixed on the A particle is considered (the corresponding tensor components will be hereafter indicated with the AB subscript), the following forms are obtained:

$$Y_B = \frac{i}{\sqrt{2}} \sum_{p=-1}^{1} \left[ D_{1p}^{1*}(\Omega_A) + D_{-1p}^{1*}(\Omega_A) \right] T_{AB}^{1p} \tag{B.2a}$$

$$Y_B^2 = -\frac{1}{\sqrt{3}} D_{00}^{0} T_{AB}^{00} + \sum_{p=-2}^{2} \left[ -\frac{1}{\sqrt{6}} D_{0p}^{2*}(\Omega_A) - \frac{1}{2} D_{2p}^{2*}(\Omega_A) - \frac{1}{2} D_{-2p}^{2*}(\Omega_A) \right] T_{AB}^{2p} \tag{B.2b}$$

where $D_{pq}^{j}$ are Wigner rotation matrices[28], $\Omega_A$ are Euler angles defining the orientation of the A particle in the laboratory frame, and the star denotes the complex conjugate.

The components $T_{AB}^{1p}$, $T_{AB}^{00}$ e $T_{AB}^{2p}$ have the following expressions in terms of spherical coordinates, $\mathbf{R}_{AB} = (R_{AB}, \phi_{AB}, \vartheta_{AB})$:

$$T_{AB}^{1\pm1} = -\left(\frac{1}{2}\right)^{\frac{1}{2}} \left(R_{AB} \sin\vartheta_{AB} \cos\phi_{AB} + \pm i R_{AB} \sin\vartheta_{AB} \sin\phi_{AB}\right) \tag{B.3a}$$

$$T_{AB}^{00} = -\frac{1}{\sqrt{3}} R_{AB}^2 \tag{B.3b}$$

$$T_{AB}^{20} = \frac{1}{\sqrt{6}} R_{AB}^2 \left(-\sin^2 \vartheta_{AB} + 2\cos^2 \vartheta_{AB}\right) \tag{B.3c}$$

$$T_{AB}^{2\pm 1} = -R_{AB} z_{AB} \sin \vartheta_{AB} \cos \vartheta_{AB} \left(\cos \phi_{AB} \pm i \sin \phi_{AB}\right) \tag{B.3d}$$

$$T_{AB}^{2\pm 2} = R_{AB}^2 \sin^2 \vartheta_{AB} \left[\frac{1}{2}\left(\cos^2 \phi_{AB} - \sin^2 \phi_{AB}\right) \pm i \cos \phi_{AB} \sin \phi_{AB}\right] \tag{B.3e}$$

## Appendix C

**Derivation of the expression for the chiral strength Eq. (31)**

If the expansions Eqs. (23) and (30a) for the density function and its first derivative are substituted into Eq. (20), and the expression for $Y_B$ in terms of the irreducible spherical components, Eq. (B.1a), is used, we obtain:

$$K_t = -\frac{i}{4\sqrt{2}} \frac{1}{(8\pi^2 v)^2} \sum_{j_A = 0,2,4,\ldots} (2j_A + 1) \langle D_{00}^{j_A} \rangle \sum_{j_B = 2,4,\ldots} (2j_B + 1) \sqrt{\frac{(j_B + 1)!}{(j_B - 1)!}} \langle D_{00}^{j_B} \rangle \times \tag{C.1}$$

$$\times \int d\mathbf{R}_B d\Omega_A d\Omega_B D_{00}^{j_A}(\Omega_A) \left[D_{10}^{j_B}(\Omega_B) - D_{-10}^{j_B}(\Omega_B)\right] \left(T_B^{11} + T_B^{1-1}\right) u(\mathbf{R}_{AB}, \Omega_{AB})$$

It is convenient to change the integration variables $R_B \rightarrow R_{AB}$, $\Omega_B \rightarrow \Omega_{AB}$ (see Chart I); then, by using the addition theorem for Wigner rotation matrices[28], Eq. (C.1) becomes:

$$K_t = -\frac{i}{4\sqrt{2}} \frac{1}{(8\pi^2 v)^2} \sum_{j_A = 0,2,4,\ldots} (2j_A + 1) \langle D_{00}^{j_A} \rangle \sum_{j_B = 2,4,\ldots} (2j_B + 1) \sqrt{\frac{(j_B + 1)!}{(j_B - 1)!}} \langle D_{00}^{j_B} \rangle \times \tag{C.2}$$

$$\times \sum_{p=-1}^{1} \sum_{r=-j_B}^{j_B} \underline{\int d\Omega_A D_{00}^{j_A}(\Omega_A) \left[D_{1r}^{j_B}(\Omega_A) - D_{-1r}^{j_B}(\Omega_A)\right] \left[D_{1,p}^{1\,*}(\Omega_A) + D_{-1,p}^{1\,*}(\Omega_A)\right]} \times$$

$$\times \int d\Omega_{AB} D_{r0}^{j_B}(\Omega_{AB}) \int d\mathbf{R}_{AB} R_{AB}^p u(\mathbf{R}_{AB}, \Omega_{AB})$$

By exploiting the orthogonality of Wigner matrices and using the coupled representation, the underlined integral can be expresses as:

$$\int d\Omega_A D_{00}^{j_A}(\Omega_A)\left[D_{1r}^{j_B}(\Omega_A)-D_{-1r}^{j_B}(\Omega_A)\right]\left[D_{1,p}^{1\ *}(\Omega_A)+D_{-1,p}^{1\ *}(\Omega_A)\right]= \quad (C.3)$$

$$=\frac{16\pi^2}{3}C^2(j_A,j_B,1;0,1,1)\delta_{p,r}\left(\delta_{p,1}-\delta_{p,-1}\right)$$

where $C(j_A, j_B, 1, 0, 1, 1)$ are Clebsch-Gordan coefficients[28] and $\delta$ is the Kronecker symbol. If Eq. (C.3) is substituted into Eq. (C.2), after some algebraic manipulation the expression Eq. (31) is obtained for the chiral strength $K_t$.

## Appendix D

**Derivation of the expression for the twist elastic constant Eqs.(32)-(33)**

If the expansions Eqs.(23) and (30b) for the density function and its second derivative are substituted into Eq.(21), we obtain:

$$K_{22}=\frac{1}{8}\left(\frac{1}{8\pi^2 v}\right)^2\sum_{j_A=0,2,4...}(2j_A+1)\langle D_{00}^{j_A}\rangle\sum_{j_B=2,4,...}(2j_B+1)\langle D_{00}^{j_B}\rangle\int d\mathbf{R}_B\int d\Omega_A\int d\Omega_B D_{00}^{j_A}(\Omega_A)\times$$

$$\times\left\{-2j_B(j_B+1)D_{00}^{j_B}(\Omega_B)+\sqrt{\frac{(j_B+2)!}{(j_B-2)!}}\left[D_{20}^{j_B}(\Omega_B)+D_{-20}^{j_B}(\Omega_B)\right]\right\}Y_B^2\,u(\mathbf{R}_{AB},\Omega_{AB})$$

(D.1)

It is convenient to change the integration variables $R_B \to R_{AB}$, $\Omega_B \to \Omega_{AB}$ (see Chart I); then, by using the addition theorem for Wigner rotation matrices[28], and the expression for $Y_B^2$ in terms of the irreducible spherical components, Eq. (B.1b), is used, Eq. (D.1) becomes:

$$K_{22} = \frac{1}{8}\left(\frac{1}{8\pi^2 v}\right)^2 \sum_{j_A=0,2,4...}(2j_A+1)\langle D_{00}^{j_A}\rangle \sum_{j_B=2,4,...}(2j_B+1)\langle D_{00}^{j_B}\rangle$$

$$\sum_{r=-j_B}^{j_B}\int d\Omega_A \int d\mathbf{R}_{AB} D_{00}^{j_A}(\Omega_A) \times$$

$$\times\left\{-2j_B(j_B+1)D_{0r}^{j_B}(\Omega_A) + \sqrt{\frac{(j_B+2)!}{(j_B-2)!}}\left[D_{2r}^{j_B}(\Omega_A) + D_{-2r}^{j_B}(\Omega_A)\right]\right\} \times \quad (D.2)$$

$$\times\left\{-\frac{2}{\sqrt{12}}D_{00}^{0*}(\Omega_A)T_{AB}^{00} + \sum_{p=-2}^{2}\left[-\frac{1}{\sqrt{6}}D_{0p}^{2*}(\Omega_A) - \frac{1}{2}D_{2p}^{2*}(\Omega_A) - \frac{1}{2}D_{-2p}^{2*}(\Omega_A)\right]T_{AB}^{2p}\right\} \times$$

$$\times \int d\Omega_{AB} D_{r0}^{j_B}(\Omega_{AB}) u(\mathbf{R}_{AB},\Omega_{AB})$$

The integrals in the $\Omega_A$ variables can be evaluated by exploiting the orthogonality of Wigner matrices and using the coupled representation. We can then write:

$$\int d\Omega_A D_{00}^{j_A}(\Omega_A)\left\{-2j_B(j_B+1)D_{0r}^{j_B}(\Omega_A) + \right.$$
$$\left. + \sqrt{\frac{(j+2)!}{(j-2)!}}\left[D_{2r}^{j_B}(\Omega_A) + D_{-2r}^{j_B}(\Omega_A)\right]\right\}\left(-\frac{2}{\sqrt{12}}D_{00}^{0*}(\Omega_A)\right) =$$
$$= \sqrt{\frac{4}{3}}8\pi^2 j_B(j_B+1) C^2(j_A,j_B,0;0,0,0)$$
(D.3a)

(D.3b)

$$\int d\Omega_A D_{00}^{j_A}(\Omega_A)\{-2j_B(j_B+1)D_{0r}^{j_B}(\Omega_A)+$$

$$+\sqrt{\frac{(j+2)!}{(j-2)!}}[D_{2r}^{j_B}(\Omega_A)+D_{-2r}^{j_B}(\Omega_A)]\}\times$$

$$\times\left(\sum_{p=-2}^{2}-\frac{1}{\sqrt{6}}D_{0p}^{2*}(\Omega_A)-\frac{1}{2}D_{2p}^{2*}(\Omega_A)-\frac{1}{2}D_{-2p}^{2*}(\Omega_A)\right)=$$

$$=\frac{8}{5}\pi^2 C(j_A,j_B,2;0,r,r)\delta_{rp}\left\{+\sqrt{\frac{2}{3}}j_B(j_B+1)C(j_A,j_B,2;0,0,0)\right.$$

$$\left.-\frac{1}{2}\sqrt{\frac{(j+2)!}{(j-2)!}}[C(j_A,j_B,2;0,2,2)+C(j_A,j_B,2;0,-2,-2)]\right\}$$

where $C(j_A,j_B,j;0,p,p)$ are Clebsch-Gordan coefficients. By substituting these espressions in Eq. (D.2) and separating the terms containing the irreducible tensors of zeroth and second rank we obtain:

$$[K_{22}]_{T^0}=\frac{1}{8}\left(\frac{1}{8\pi^2 v}\right)^2 \sum_{j_A=0,2,4...}(2j_A+1)\langle D_{00}^{j_A}\rangle \sum_{j_B=2,4,...}(2j_B+1)\langle D_{00}^{j_B}\rangle \int d\mathbf{R}_{AB}\int d\Omega_{AB}\delta_{j_A j_B}\times$$

$$\times\left\{-2j_B(j_B+1)C^2(j_A,j_B,0;0,0,0)8\pi^2 D_{00}^{j_B}(\Omega_{AB})\left(-\frac{2}{\sqrt{12}}\right)T_{AB}^{00}u(\mathbf{R}_{AB},\Omega_{AB})\right. \quad\text{(D.4a)}$$

$$[K_{22}]_{T^2}=\frac{1}{8}\left(\frac{1}{8\pi^2 v}\right)^2 \sum_{j_A=0,2,4...}(2j_A+1)\langle D_{00}^{j_A}\rangle \sum_{j_B=2,4,...}(2j_B+1)\langle D_{00}^{j_B}\rangle \int d\mathbf{R}_{AB}\int d\Omega_{AB}\times$$

(D.4b)
$$\times\left\{\sum_{p=-2}^{2}\frac{8\pi^2}{5}\left[-2j_B(j_B+1)C(j_A,j_B,2;0,0,0)C(j_A,j_B,2;0,p,p)\left(-\frac{1}{\sqrt{6}}\right)-\right.\right.$$

$$\left.\left.-\sqrt{\frac{(j_B+2)!}{(j_B-2)!}}C(j_A,j_B,2;0,2,2)C(j_A,j_B,2;0,p,p)\right]D_{p0}^{j_B}(\Omega_{AB})T_{AB}^{2p}\right\}u(\mathbf{R}_{AB},\Omega_{AB})$$

If the summation over the *p* index is made explicit and the terms with *p*=1,-1, *p*=2,-2 are collected, Eqs. (33) are obtained.

# Appendix E

**Demonstration of the vanishing chiral electrostatic contribution when averaged over a cylindrically symmetric pair distribution**

A generic integral in the electrostatic part of Eq. (31) can be written as:

$$\int_0^\infty dR_{AB} R_{AB}^3 \int_0^\pi d\vartheta_{AB} \sin\vartheta_{AB} \int_0^\pi d\beta_{AB} \sin\beta_{AB} d_{10}^j(\beta_{AB}) \int_0^{2\pi} d\phi_{AB} \int_0^{2\pi} d\alpha_{AB} \sin(\alpha_{AB} - \phi_{AB})$$
$$\int_0^{2\pi} d\gamma_{AB}\, g_h(R_{AB},\phi_{AB},\vartheta_{AB},\alpha_{AB},\beta_{AB},\gamma_{AB}) U_{el}(R_{AB},\phi_{AB},\vartheta_{AB},\alpha_{AB},\beta_{AB},\gamma_{AB})$$
(E.1)

For cylinders the correlation function $g_h$ is independent of the angle $\gamma_{AB}$, so we can write:

$$\int_0^\infty dR_{AB} R_{AB}^3 \int_0^\pi d\vartheta_{AB} \sin\vartheta_{AB} \int_0^\pi d\beta_{AB} \sin\beta_{AB} d_{10}^j(\beta_{AB}) \int_0^{2\pi} d\phi_{AB} \int_0^{2\pi} d\alpha_{AB} \sin(\alpha_{AB} - \phi_{AB}) \times$$
$$\times g_h(R_{AB},\phi_{AB},Z_{AB},\alpha_{AB},\beta_{AB}) \overline{U_{el}}(R_{AB},\phi_{AB},Z_{AB},\alpha_{AB},\beta_{AB})$$
(E.2)

where $\overline{U_{el}}(R_{AB},\phi_{AB},\vartheta_{AB},\alpha_{AB},\beta_{AB})$ represents the electrostatic interaction between charges on molecule A and charged rings on molecule B. After a change in the integration variables we obtain:

$$\int_0^\infty dR_{AB} R_{AB}^3 \int_0^\pi d\vartheta_{AB} \sin\vartheta_{AB} \int_0^\pi d\beta_{AB} \sin\beta_{AB} d_{10}^j(\beta_{AB}) \int_0^{2\pi} d\phi_{AB} \int_0^{2\pi} d\alpha'_{AB} \sin(\alpha'_{AB}) \times$$
$$\times g_h(\alpha'_{AB}+\phi_{AB},\beta_{AB},R_{AB},\vartheta_{AB},\phi_{AB}) \overline{U_{el}}(\alpha'_{AB}+\phi_{AB},\beta_{AB},R_{AB},\vartheta_{AB},\phi_{AB})$$
(E.3)

which can be rearranged as:

$$\int_0^\infty dR_{AB} R_{AB}^3 \int_0^\pi d\vartheta_{AB} \sin\vartheta_{AB} \int_0^\pi d\beta_{AB} \sin\beta_{AB} d_{10}^j(\beta_{AB}) \int_0^\pi d\alpha'_{AB} \sin(\alpha'_{AB}) \int_0^{2\pi} d\phi_{AB} \times$$
$$\times [\, g_h(\alpha'_{AB}+\phi_{AB},\beta_{AB},R_{AB},\vartheta_{AB},\phi_{AB}) \overline{U_{el}}(\alpha'_{AB}+\phi_{AB},\beta_{AB},R_{AB},\vartheta_{AB},\phi_{AB}) -$$
$$- g_h(\alpha'_{AB}+\phi_{AB}+\pi,\beta_{AB},R_{AB},\vartheta_{AB},\phi_{AB}) \overline{U_{el}}(\alpha'_{AB}+\phi_{AB}+\pi,\beta_{AB},R_{AB},\vartheta_{AB},\phi_{AB}) \,]$$
(E.4)

The correlation function $g_h$ is independent of the $\phi_{AB}$ angle for two cylinders. Therefore integration over this variable is equivalent to calculation of the interaction between charged rings on molecule A with charged rings on molecule B. As a consequence the two integrals over the angle $\phi_{AB}$ are equal and their sum vanishes, for any value of the variables $\alpha'_{AB}, \beta_{AB}, R_{AB}, \vartheta_{AB}$.

## Appendix F

**Algorithm for the calculation of the closest approach distance between polyelectrolytes**

For given values of the variables $\alpha_{AB}, \beta_{AB}, \gamma_{AB}, \vartheta_{AB}, \phi_{AB}$, for each pair of spheres (*i,j*) the following calculations are performed:

1. the distance between the two spheres ($r_{ij}$) is expressed in terms of the intermolecular vector $\mathbf{R}_{AB}$;

2. condition of contact is imposed: $r_{ij} = \sigma_{ij}$, with $\sigma_{ij} = (\sigma_i + \sigma_j)/2$, and the closest approach distance for the two spheres, $R_{AB}^{ij}$, is stored.

At the end of the cycle the largest value of $R_{AB}^{ij}$ is taken as the closest approach distance $R_{AB}^0$.

---

## Appendix G

**Minimization of the free energy**

In practice, it is convenient to choose as variational parameters, in place of the order parameters, the expansion coefficients of the mean field potential $U^{mf}(\Omega_B)$. This is derived by imposing the stationarity condition to the free energy density:

$$\frac{\delta}{\delta \rho(\Omega_B)} \left\{ \frac{f[\rho]}{k_B T} - \lambda \int d\Omega_B \rho(\Omega_B) \right\} = 0 \tag{G.1}$$

where $\lambda$ is a Lagrange undetermined multiplier, introduced to take into account the normalization condition on the density function, Eq. (11). By calculating the functional derivative with the expressions for the free energy density Eqs. (34), (35) we can write:

$$\rho(\Omega_B) = \frac{\exp\left\{-\frac{U^{mf}(\Omega_B)}{k_B T}\right\}}{\int d\Omega_B \exp\left\{-\frac{U^{mf}(\Omega_B)}{k_B T}\right\}} \tag{G.2}$$

with the mean field potential

$$U^{mf}(\Omega_B) = \int d\mathbf{R}_{AB}\, d\Omega_A\, \rho(\Omega_A) u(\mathbf{R}_{AB},\Omega_{AB}) \tag{G.3}$$

This can be expressed by an expansion on a basis of Wigner rotation matrices:

$$\frac{U^{mf}(\Omega_B)}{k_B T} = \sum_{j=2,4..} u_j D_{00}^j(\Omega_B) \tag{G.4}$$

After substitution of this expression into eq. (G.2) we can rewrite the free energy density of the undeformed nematic phase, eqs. (34)-(35), in the form

$$\begin{aligned}\frac{f(\{u_j\})}{k_B T} &= -\frac{1}{v}\sum_{j=2,4..} u_j \langle D_{00}^j \rangle - \frac{1}{v}(\ln v - \ln Z) + \\ &+ \frac{1}{16\pi^2 v^2}\sum_{j_A=0,2,4,...}(2j_A+1)\langle D_{00}^{j_A} \rangle^2 \int d\mathbf{R}_{AB}\, d\Omega_{AB} D_{00}^{j_A}(\Omega_{AB})\frac{u(\mathbf{R}_{AB},\Omega_{AB})}{k_B T}\end{aligned} \tag{G.5}$$

with the partition function

$$Z = \int d\Omega_B \exp\left\{-\sum_j u_j D_{00}^j(\Omega_B)\right\} \tag{G.6}$$

and the order parameters

$$\langle D_{00}^j \rangle = \frac{\int d\Omega_B \exp\left\{-\sum_j u_j D_{00}^j(\Omega_B)\right\} D_{00}^j(\Omega_B)}{Z} \tag{G.7}$$

The free energy Eq. (G.5) is thus expressed as a function of the coefficients $\{u_j\}$, and its equilibrium value can be obtained by minimizing with respect to these parameters. The use of the coefficients $\{u_j\}$ as variational parameters has a twofold advantage: (i) the expansion of the mean field potential Eq. (G.4) converges faster than that of the density function Eq. (23); (ii) the parameters $\{u_j\}$ are unconstrained, unlike order parameters, which are restricted to the range $-1/2 \leq \langle D_{00}^j \rangle \leq 1$.

# Figures

**Figure 1** Reference frames and transformations.

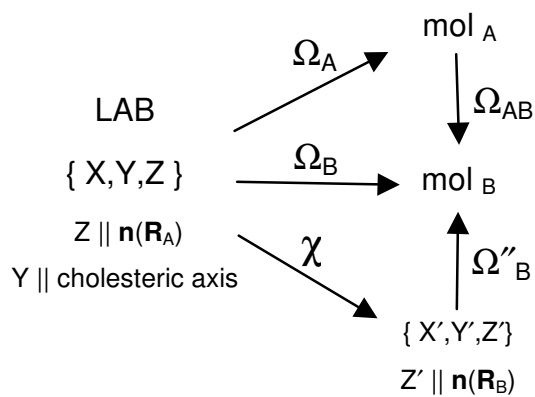

**Figure 2** Definition of the twist angle $\chi$. Black arrows indicate the orientation of the director $\hat{n}$ at different positions along cholesteric axis (Y).

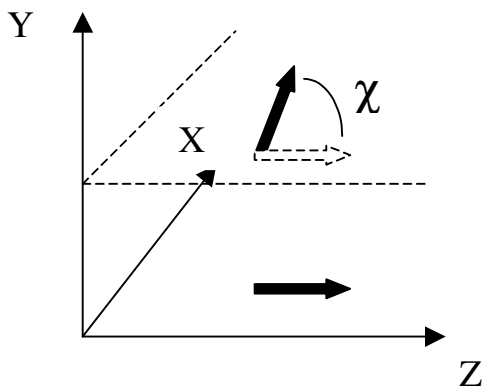

**Figure 3** Geometry and charges assumed for a B-DNA base pair. Angles are expressed in degrees and lengths in nanometers.

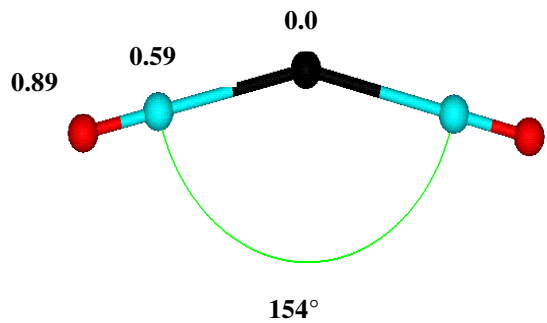

**Figure 4** Model of a B-DNA fragment of about 30 base pairs.

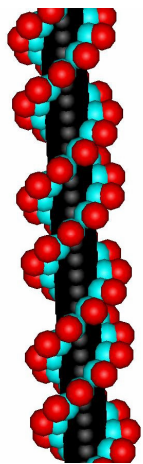

**Figure 5** A pair configuration with B-DNA molecules at close distance.

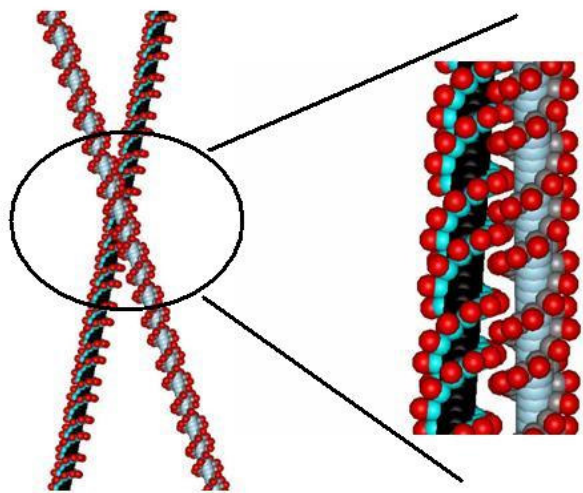

**Figure 6** Predicted cholesteric wavenumber as a function of the inverse temperature (solid line). Steric contribution and experimental results [9] are shown by the dot-dashed and the dashed line, respectively.

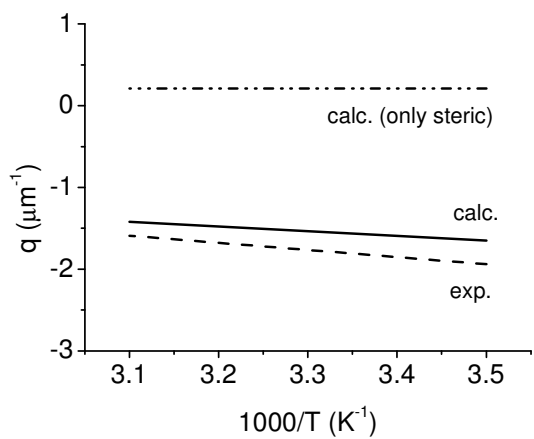

**Table I** Parameters for the spheres used for modeling the shape of B-DNA.

| group | sphere diameter $\sigma$ (nm) | distance of sphere center from helical axis (nm) |
|---|---|---|
| base-pairs | 1.0 | 0.0 |
| sugar | 0.6 | 0.59 |
| phosphate | 0.6 | 0.89 |

**Table II** Characteristics of B-DNA solutions considered in our calculations.

| molecular weight (Dalton/130 bp) | molecular volume $V_0$ (nm$^3$) | concentration mg/ml | concentration mol/l | solution volume for molecule $V$ (nm$^3$) |
|---|---|---|---|---|
| $8.45 \times 10^4$ | $1.64 \times 10^2$ | 200 | $2.37 \times 10^{-3}$ | $7.0 \times 10^2$ |

**Table III** Values of Debye screening length $\kappa_D^{-1}$, dielectric constant $\varepsilon$ and fraction of uncompensated charge fraction $\delta$ employed for calculations, at different temperatures and ionic strengths.

|  | T= 286 K | | | T= 323 K | | |
|---|---|---|---|---|---|---|
| I (mol/l) | 0.1 | 0.2 | 0.5 | 0.1 | 0.2 | 0.5 |
| $\kappa_D^{-1}$ (nm) | 0.94 | 0.66 | 0.42 | 0.93 | 0.66 | 0.42 |
| $\varepsilon$ [44] | 82 | 82 | 82 | 70 | 70 | 70 |
| $\delta$ | 0.24 | 0.24 | 0.24 | 0.23 | 0.23 | 0.23 |

**Table IV.** Order parameters calculated for the system of hard helices.

| $\langle D_{00}^2 \rangle$ | 0.89 |
|---|---|
| $\langle D_{00}^4 \rangle$ | 0.68 |
| $\langle D_{00}^6 \rangle$ | 0.43 |

**Table V.** Elastic constant, chiral strength and cholesteric wavenumber predicted for a system of hard helices.

| $K_{22}$ ($10^{-12}$ N) | $K_t$ ($10^{-6}$ N/m) | $q$ (μm$^{-1}$) |
|---|---|---|
| 1.00 | -0.21 | 0.21 |

**Table VI.** Order parameters predicted for B-DNA solutions at different temperatures and ionic strengths.

|  | T=286 K | T=323 K | | |
|---|---|---|---|---|
| I (mol/l) | 0.2 | 0.1 | 0.2 | 0.5 |
| $\langle D_{00}^2 \rangle$ | 0.87 | 0.83 | 0.86 | 0.88 |
| $\langle D_{00}^4 \rangle$ | 0.62 | 0.55 | 0.63 | 0.65 |
| $\langle D_{00}^6 \rangle$ | 0.38 | 0.30 | 0.39 | 0.41 |

**Table VII** Electrostatic contributions to elastic constants and chiral strengths (the steric contributions are reported in Table V), along with cholesteric wavenumbers predicted for B-DNA solutions at different temperatures and ionic strengths.

|  | T=286 K | T=323 K | | |
|---|---|---|---|---|
| I (mol/l) | 0.2 | 0.1 | 0.2 | 0.5 |
| $K_{22}$ ($10^{-12}$ N)    electr. | -0.16 | -0.16 | -0.09 | -0.01 |
| $K_t$ ($10^{-6}$ N/m)    electr. | 1.79 | 1.71 | 1.50 | 1.41 |
| $q$ (µm$^{-1}$) | -1.65 | -1.80 | -1.42 | -1.20 |

**Table VIII.** Contributions of different rank to twist elastic constants; steric and electrostatic parts are reported. Values are calculated for the temperature T=323K and the ionic strength I=0.2 mol/l.

| $K_{22}^{j_A j_B}$ ($10^{-12}$ N) | steric | electrostatic |
|---|---|---|
| $K_{22}^{22}$ | 2.27 | -0.21 |
| $K_{22}^{24}$ | -0.95 | 0.13 |
| $K_{22}^{42}$ | -0.95 | 0.13 |
| $K_{22}^{44}$ | 0.88 | -0.19 |
| $K_{22}^{46}$ | -0.29 | 0.08 |
| $K_{22}^{64}$ | -0.30 | 0.08 |
| $K_{22}^{66}$ | 0.34 | -0.11 |

**Table IX.** Contributions of different rank to chiral strength; steric and electrostatic parts are reported. Values are calculated for the temperature T=323K and the ionic strength I=0.2 mol/l.

| $K_t^{j_A j_A}$ ($10^{-6}$ N/m) | steric | electrostatic |
|---|---|---|
| $K_t^{22}$ | -0.34 | 1.79 |
| $K_t^{44}$ | 0.15 | -0.38 |
| $K_t^{66}$ | -0.02 | 0.11 |

**Table X.** Effect of truncation at different ranks on elastic constant, chiral strangth and cholesteric wavenumber, calculated at the temperature T=323 K and ionic strength I=0.2 mol/l.

| rank | | 2 | 4 | 6 |
|---|---|---|---|---|
| $K_{22}$ | steric | 2.27 | 1.25 | 1.00 |
| $(10^{-12}\,N)$ | electrostatic | -0.21 | -0.14 | -0.1 |
| $K_t$ | steric | -0.34 | -0.19 | -0.21 |
| $(10^{-6}\,N/m)$ | electrostatic | 1.79 | 1.39 | 1.50 |
| q ($\mu m^{-1}$) | | -0.70 | -1.08 | -1.42 |